\documentclass[twocolumn,showpacs,prb,10pt]{revtex4}
\usepackage{graphicx}% Include figure files
\usepackage{dcolumn}% Align table columns on decimal point
\usepackage{bm}% bold math

\begin{document}
\title{White lines and $d$ occupancy for the 3$d$ transition-metal oxides and lithium transition-metal oxides}\author{Jason Graetz,
Channing C. Ahn, Hao Ouyang$^1$, Peter Rez$^2$, and Brent Fultz}
\affiliation{Division Engineering and Applied Science M/C 138-78, California Institute of Technology, Pasadena, California 91125
\\$^1$National Chung Hsing University, Department of Materials Engineering, Taichung, Taiwan 402, R.O.C.
\\$^2$Department of Physics and Astronomy and CSSS, Arizona State University, Tempe, AZ 85287-1504}
\date{\today}
\pacs{79.20.Uv,71.20.Be,82.80.Pv,82.47.Aa}
\baselineskip=10pt
\begin{abstract}
Electron energy-loss spectrometry was employed to measure the white lines at the $L_{2,3}$ absorption edges of the 3$d$ transition-metal oxides and lithium
transition-metal oxides. The white-line ratio ($L_3/L_2$) was found to increase between $d^0$ and $d^5$ and decrease between $d^5$ and
$d^{10}$, consistent with previous results for the transition metals and their oxides. The intensities of the white lines, normalized to the post-edge
background, are linear for the 3$d$ transition-metal oxides and lithium transition-metal oxides. An empirical correlation between normalized white-line intensity
and 3$d$ occupancy is established. It provides a method for measuring changes in the 3$d$-state occupancy. As an example, this empirical relationship is used to
measure changes in the transition-metal valences of Li$_{1-x}$Ni$_{0.8}$Co$_{0.2}$O$_2$ in the range of $0 \leq x \leq 0.64$. In these experiments the 3$d$
occupancy of the nickel ion decreased upon lithium deintercalation, while the cobalt valence remained constant.
\end{abstract}
\maketitle

\section{Introduction}
Transition-metal oxides are one of the most fascinating classes of inorganic solids, exhibiting a wide variety of structures, properties, and phenomena. 
The unusual properties of the transition-metal oxides are typically attributed to the unique nature of the outer $d$ electrons \cite{Rao_book}. In this study, the
outer electron shells of the $3d$ transition-metal oxides and lithium transition-metal oxides are characterized using electron energy-loss spectrometry
(EELS). The $3d$ transition-metal oxides are well suited to this technique due to their low energy core-edges ($<1$ keV) and their inherent stability under
the electron beam. The near-edge structure of the EELS core-edges are especially interesting because they reflect a density of unoccupied states of an atom or
ion, which can be used to characterize valences and atomic bonding.

A method for measuring $d$-state occupancy is valuable for understanding the unique character of the transition-metal oxides and lithiated transition-metal
oxides. The discovery of high temperature superconductivity in perovskites has generated tremendous interest in the electronic structure of the transition-metal
oxides. Recent advances in oxide-based spintronic materials and the discovery of colossal magnetoresistance in manganese perovskites has created renewed interest
in the unique electronic environments of the transition-metal ion and the role of charge ordering in these materials. Other research has been driven by the
success of the lithiated transition-metal oxides as cathodes in rechargeable lithium batteries. A number of x-ray absorption and EELS experiments have been
performed to investigate charge compensation during electrochemical cycling of lithium
\cite{Saadoune,Delmas,Balasubramanian,Mansour1,Mansour2,Mansour3,Nakai1,Nakai2,Nakai3,Nonaka,Montoro2000,Montoronico,Uchimoto,Graetz2002,Yoon,Graetzlinicoo2,Xu2003,Chen2003}.
However, these studies are predominately qualitative. We have performed a quantitative investigation of the transition-metal valences in 
Li$_{1-x}$Ni$_{0.8}$Co$_{0.2}$O$_{2}$ to determine the exact change in $d$-state occupancy between $x = 0$ and $x = 0.64$.

LiNi$_y$Co$_{1-y}$O$_2$ has a rhombohedral symmetry (R$\bar{3}$m space group), where the transition-metal ions are present in a solid solution and are
octahedrally coordinated by oxygen ions. In LiCoO$_2$ and LiNi$_y$Co$_{1-y}$O$_2$ the oxygen ion accommodates much of the Li 2$s$ electron during lithiation
\cite{Montoro2000,Montoronico,Graetz2002,Graetzlinicoo2}. In LiNi$_y$Co$_{1-y}$O$_2$ both the nickel and oxygen ions are believed to participate in charge
compensation around the Li$^+$ ions \cite{Graetzlinicoo2} and it is unclear how much charge is transferred to each transition-metal ion. It is our contention
that changes in the transition-metal valence can be measured quantitatively using the transition-metal white lines.

The $L_{2,3}$-edge is composed of two, independent, overlapping $L_2$ and $L_3$-edges, resulting from transitions from $2p$
states into bound $3d$ and continuum states. The splitting of the $L$-edge is due to a spin-orbit coupling that breaks the degeneracy of the $2p$ states into
$2p_{1/2}$ and $2p_{3/2}$ levels. The transition-metal $L_2$- and $L_3$-edges are each characterized by an intense peak at the edge onset, known as a ``white
line''. This near-edge structure is due to transitions into the highly localized 3$d$ states near the Fermi level. Many efforts based on the one-electron
approximation or band-structure studies have tried to relate the variation of the total white-line intensities observed during alloying to the change in the
filling of the corresponding outer $d$ states \cite{Mattheiss,Mansour1984,Pearson2}.  Since the 2$p$ states are atomic and 3$d$ states are tightly bound about a
given atom in a solid, the information obtained from the white lines is largely local to a given atom species \cite{Rez_EELS}. Pearson et
al.~\cite{Pearson1,Pearson2} found a linear correlation between white-line intensity and $d$ occupancy for the $3d$ and $4d$ transition metals. In this work, we
extend the former effort of Pearson et al.~\cite{Pearson1,Pearson2} from 3$d$ transition metals to 3$d$ transition-metal oxides and some lithium based oxides.

The onset energies of the transition-metal $L_{2,3}$-edges are sensitive to the oxidation state of the transition-metal ion \cite{Tafto,Leapman}. 
For example, the Cu $L_{3}$-edge has been shown to shift from 933 eV to 931 eV between Cu and CuO (Cu$^{0}$ to Cu$^{2+}$) \cite{Leapman}. Another method for
extracting valence information from quantified energy-loss spectra is through the white-line ratio ($L_3/L_2$). When the initial and final states are uncoupled,
the white-line intensity ratio has a statistical value of 2. The $p$ states have a degeneracy of $2j+1$. Therefore, there are twice as many $j=3/2$ ($L_{3}$)
electrons as there are $j=1/2$ ($L_{2}$) electrons. Deviations from this value were originally reported by Leapman and Grunes for the early 3$d$
transition metals \cite{Leapman_ratios} and have also been reported in other transition metal and lanthanide compounds
\cite{Leapman,Sparrow,Leapman_ratios,Thole1988,Wendin,Thole1985,Rao}. These deviations depend on the configuration of the outer electron shell, and are difficult
to interpret without detailed information on the local chemical environment.

A simpler and more direct approach to characterizing the transition-metal 3$d$ states is to measure the integrated intensity of the white lines. The white-lines
are representative of the component of the 3$d$ band locally projected onto the 3$d$ atomic states at the transition-metal atoms. The white-line
intensity is a function of the number of $2p \rightarrow 3d$ excitations. The probability of an excitation from an initial state, $i$, to a final state,
$f$, depends on the density of unoccupied final states, $\rho_{f}^{u}(E)$,
\begin{equation}
I \propto \frac{1}{q^4}\rho_{f}^{u}(E)|\langle f | e^{i\mbox{\boldmath $q$} \cdot \mbox{\boldmath $r$}} | i \rangle|^2,
\end{equation}
where $\mbox{\boldmath $q$}$ and $\mbox{\boldmath $r$}$ are the momentum transfer and position vectors, respectively. The white-line intensity is a direct measure
of the number of unoccupied $3d$ states, or holes, in the 3$d$ electron band \cite{Starace,Horsley1982,Rao}. An inverse linear relationship between the integrated
white-line intensity and the atomic number ($d$ occupancy) is well documented for the $3d$ transition metals
\cite{Pearson1,Pearson2}. By performing measurements of  white-line intensities on oxides of transition metals across the periodic table, we tested the quality
of an empirical procedure for obtaining the $d$-electron occupancy from the intensity of the white lines. The results indicate that it is practical to use white
lines to measure changes in $d$-electron occupancy in similar oxide materials. The utility of this relationship is demonstrated for the lithiation of
LiNi$_{0.8}$Co$_{0.2}$O$_2$.

\section{Experimental}
The transition-metal oxide samples were prepared by thermal evaporation on single-crystal NaCl substrates in vacuum ($10^{-5}$ torr). The samples were
annealed in air at 350$^{\circ}$ C for two hours to oxidize the films and subsequently floated onto a holey carbon grid in water. The lithium transition-metal
oxide powders were crushed in Fluorinert$^{\textup{\scriptsize{TM}}}$ with a mortar and pestle and the particles were floated onto a holey carbon grid. The
transition-metal oxide spectra (excluding CrO$_x$) were collected using a Gatan 607 serial-detection spectrometer on a Jeol-200CX TEM operating at 200 kV. The
spectra for the chromium oxide and the lithiated compounds were acquired using a Gatan 666 parallel-detection spectrometer on a Philips EM420 at 100 kV. The
low-loss and core-loss spectra were acquired separately to maximize the energy resolution and detective quantum efficiency. The energy resolution was
approximately 1.5 eV for both spectrometers.

Li$_{0.36}$Ni$_{0.8}$Co$_{0.2}$O$_2$ was prepared by delithiating the
stoichiometric material (Li$_{1.0}$Ni$_{0.8}$Co$_{0.2}$O$_2$) using an aqueous solution of potassium persulfate (K$_2$S$_2$O$_8$). The delithiated material was
dried in air at approximately $60^{\circ}$ C for 24 hours. A compositional analysis was performed using an inductively-coupled plasma mass spectrometer.

\begin{table}
\caption{\label{tab:table1}Composition and $d$ occupancy of the transition-metal oxide samples.}
\begin{ruledtabular}
\begin{tabular}{ccc}
Sample & x in TMO$_x$ & 3$d$ Occupancy\\
\hline
ScO$_x$ & 1.79 & 0.00\\
TiO$_x$ & 2.22 & 0.00\\
CrO$_x$ & 1.13 & 3.74\\
MnO$_x$ & 1.19 & 4.64\\
FeO$_x$ & 1.32 & 5.36\\
CoO$_x$ & 1.25 & 6.50\\
NiO$_x$ & 1.00 & 8.00\\
CuO$_x$ & 0.93 & 9.14\label{table}
\end{tabular}
\end{ruledtabular}
\end{table}
The lithiated samples were assumed to be stoichiometric (except for the delithiated Li$_{1-x}$Ni$_{0.8}$Co$_{0.2}$O$_2$). The composition of
the transition-metal oxides ($x$ in TMO$_x$) was determined through an elemental analysis of the transition metal and oxygen core-edges
valid for thin films ($<100$ nm),
\begin{equation}
\frac{N_{\textup{\scriptsize{O}}}}{N_{\textup{\scriptsize{TM}}}}=
\frac{I_{\textup{\scriptsize{O}} K}}{I_{\textup{\scriptsize{TM}} L}}\frac{\sigma_{\textup{\scriptsize{TM}} L}(\Delta)}{\sigma_{\textup{\scriptsize{O}}
K}(\Delta)},
\end{equation}
where $N$ is the number of atoms, $I$ is the intensity of the ionization edge over an integration window $\Delta$, and
$\sigma(\Delta)$ is the reduced cross section. In most cases, electron diffraction patterns showed that the metal oxide films were non-stoichiometric, single
phases. The $d$ occupancy of each sample was determined by the oxidation state of the transition metal. In these oxides, the 3$d$ electrons are typically
involved in bonding with the oxygen ion (specifically the O 2$p$ states). However, the exact amount of $d$-character in these materials is unknown. For
simplicity, and to be consistent with the results for the 3$d$ transition metals \cite{Pearson1,Pearson2}, the effects due to hybridization were ignored. This
approximation should not alter the trend across the series of 3$d$ transition metal elements. It was assumed that each oxygen atom removed (or oxidized) two
electrons from each transition-metal ion, and the transition-metal ions relinquished their $4s$ electrons first, followed by their $3d$ electrons. In the
elemental state (prior to oxidation) each of the transition metals was assumed to have a 3$d^{n-2}s^2$ valence-electron configuration, where $n$ is the number of
valence electrons, except for chromium and copper, which have a 3$d^{n-1}s^1$ configuration. In this method, the $3d$ occupancies of scandium oxide and titanium
oxide are negative, suggesting that the oxidation extends into the $3p$ band. The $3d$ occupancy of these oxides was assumed to be zero. A list of the
composition and 3$d$ occupancy of the transition-metal oxides are presented in table
\ref{table}.

\section{Data Analysis}
Steps were taken to reduce artifacts in the measured spectra. Each spectrum was divided by a gain calibration
spectrum, or instrument response, to account for the variation in electron response due to nonuniformities in the scintillator plate and the unique photonic
sensitivity of each diode of the EELS detector. An instrument response was acquired by applying parallel illumination to the entire detector array and normalizing
to the integrated intensity. Gain averaging was employed to reduce channel-to-channel gain fluctuations \cite{Shuman}.  The plural scattering was
removed using a Fourier-log deconvolution \cite{Egerton} and the background was approximated and subtracted from the core-edge. The post-edge region was
characterized by a decay governed by the ionization cross section. The background was therefore approximated as a power-law of the form $I(E) = AE^{-r}$, where
$E$ is the energy loss, and $A$ and $r$ are constants obtained by fitting the pre-edge background.

\begin{figure}
\includegraphics{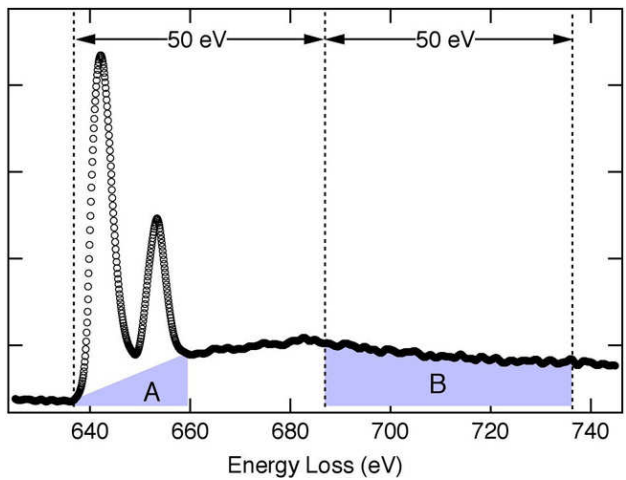}
\caption{\label{L23_windows}(Color online) Mn $L_{2,3}$-edge from MnO$_x$ showing the free-electron-like contributions to the white-line intensity (Region A) and
a 50 eV window located 50 eV above the $L_3$ onset (Region B), which was used to normalize the white-line intensity.}
\end{figure}

The spectral intensity was partitioned into contributions from bound state transitions (forming the white lines) and
transitions into delocalized states. The intensity due to continuum (free-electron-like) transitions was removed to restrict the analysis to bound
$3d$ transitions. The free-electron-like contributions to the core-edge were approximated as the total intensity below the Gaussian peaks (Figure
\ref{L23_windows} window A).
To quantify the white-line
intensity, the continuum component of the near-edge structure was removed and the total $L_{2,3}$ white-line intensity was normalized to a 50
eV continuum window, 50 eV above the $L_3$-edge onset (Figure
\ref{L23_windows} window B). The spectral intensity from a transition of a bound electron in a $|2p
\rangle$ state to an excited $|3d \rangle$ state is the product of the inelastic form factor governing the transition and the density of unoccupied states
$\rho^{u} (3d)$. According to Mattheiss and Deitz \cite{Mattheiss} and Pearson et al.~ \cite{Pearson2} for an atomic, one-electron model, this relationship can
be approximated as:
\begin{equation}
I \propto \rho^{u}(3d)|\langle 3d|\exp[i\mbox{\boldmath $q$} \cdot \mbox{\boldmath $r$}]|2p \rangle |^2.
\end{equation}
The total white-line intensity is the sum of all the transitions within a given energy region. In this experiment the white-line intensity
($E_0 \leq E \leq E_1$) has been normalized to the continuum to give the following expression for the measured
intensity:
\begin{displaymath}
I \approx \rho^u(3d)\frac{\int_{E_0}^{E_1}|\langle 3d|\exp[i\mbox{\boldmath $q$} \cdot \mbox{\boldmath $r$}]|2p \rangle |^2 \textup{d}
\varepsilon}{\int_{E_0+50}^{E_0+100}|
\langle 3d| \exp[i\mbox{\boldmath $q$} \cdot \mbox{\boldmath $r$}] |2p \rangle |^2 \textup{d} \varepsilon}
\end{displaymath}
\begin{equation}= \rho^u(3d)\frac{M_{\textup{white
line}}}{M_{\textup{continuum}}},
\label{eq4}
\end{equation}
\noindent where $E_0$ is the edge onset energy and $E_1$ is the energy at the trailing edge of the $L_2$ line, typically $E_1 \approx E_0+30$ eV.

The matrix elements of equation \ref{eq4} were calculated using a one electron, Hartree-Slater wave
function generated by the computer code of Herman and Skillman \cite{Herman}. The ratio of these matrix elements ($M_{\textup{white
line}}/M_{\textup{continuum}}$) for elemental Sc and Mn are 0.075 and 0.148,
respectively. The  values for Ti, V, Cr, Fe, Co, Ni, and Cu were published in an earlier work
\cite{Pearson2}. The intensities of the white lines, normalized by the post-edge background, can reflect the filling of the $d$ states and were used to monitor
the change of 3$d$ occupancies while alloying.

\begin{figure}
\includegraphics{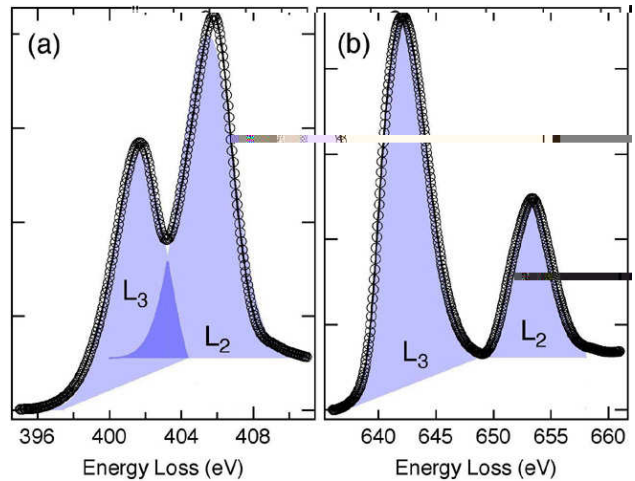}
\caption{\label{ratio_windows}(Color online) ({\bf a}) Sc $L_{2,3}$-edge from ScO$_x$ showing the typical Gaussian fit and integration windows (shaded)
used to determine the
$L_3/L_2$ ratio in the early transition metals and Cu. ({\bf b}) Mn $L_{2,3}$-edge from MnO$_x$ showing the typical integration windows (shaded) used to
determine the
$L_3/L_2$ ratio in the middle transition metals.}
\end{figure}

The white-line ratio was determined by removing the continuum component from the $L$-edge after background subtraction, and taking the ratio of the integrated
intensities of the $L_3$ and $L_2$ white lines. Due to the overlap of the $L_3$ and $L_2$ peaks in the early transition metals (Sc, Ti, and Cr), the white lines
were approximated as Gaussian functions (Figure \ref{ratio_windows}a). Similarly, Gaussian functions were also used to determine the Cu $L_3/L_2$ ratio due to the
broad near-edge structure at the Cu $L$-edge. For the middle transition metals (Mn, Fe, Co, and Ni), for which the white lines are sharp and well separated, the
ratio was determined by directly integrating the individual peaks (Figure \ref{ratio_windows}b).

\section{Results and Discussion}
\subsection{White-Line Ratios}
\begin{figure}
\includegraphics{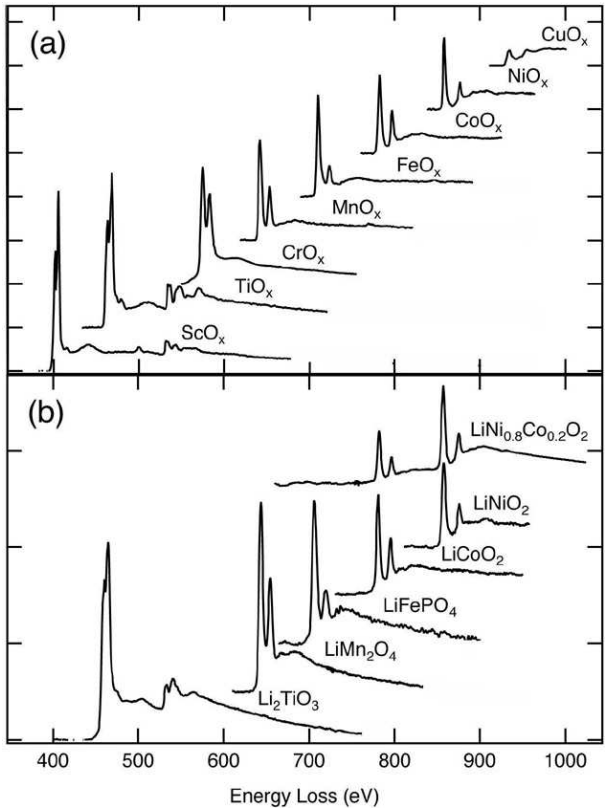}
\caption{Energy-loss spectra of the transition-metal $L_{2,3}$-edges from ({\bf a}) a series of transition-metal
oxides and ({\bf b}) lithium transition-metal oxides.}
\label{TMO_whitelines}
\end{figure}

\begin{figure}
\includegraphics{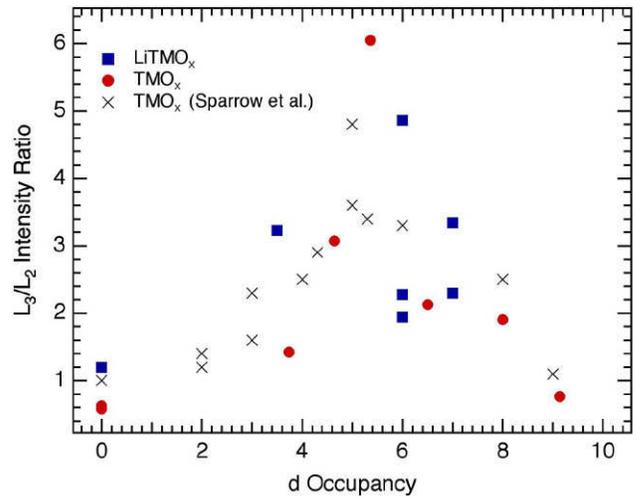}
\caption{(Color online) White-line ratio for a series of transition-metal oxides and lithium transition-metal oxides. The data of Sparrow et al.~are from
reference
\cite{Sparrow}.}
\label{branch_ratio}
\end{figure}

The transition-metal $L_{2,3}$-edges from a number of transition-metal oxides and lithium transition-metal oxides are displayed in Figure~\ref{TMO_whitelines}.
Some qualitative trends are apparent in this plot. The total white-line intensity clearly decreases with the atomic number of the transition metal. Similarly,
changes in the white-line ratios are also evident in these spectra, especially between the Ti and Fe compounds. The nonstatistical value of the white-line ratio
(i.e., a departure from 2) has been shown to vary with the
$d$ occupancy across the $3d$ transition-metal series \cite{Leapman,Sparrow,Leapman_ratios,Thole1988}. Figure \ref{branch_ratio} shows an increasing $L_3/L_2$
ratio with the filling of the $3d$ states up to approximately FeO$_x$ ($d^{5.3}$) where $L_3/L_2 \approx 6.0$.
Subsequent filling of the $d$ states reduces the white-line ratio. This is consistent with results of Sparrow et al.~who have found that in transition-metal
oxides, the white-line ratio is at a minimum in the $d^0$ and $d^9$ configurations and reaches a maximum of 4.8 when the $d$ states are half full \cite{Sparrow}.

\subsection{White-Line Intensity}
\begin{figure}[b]
\includegraphics{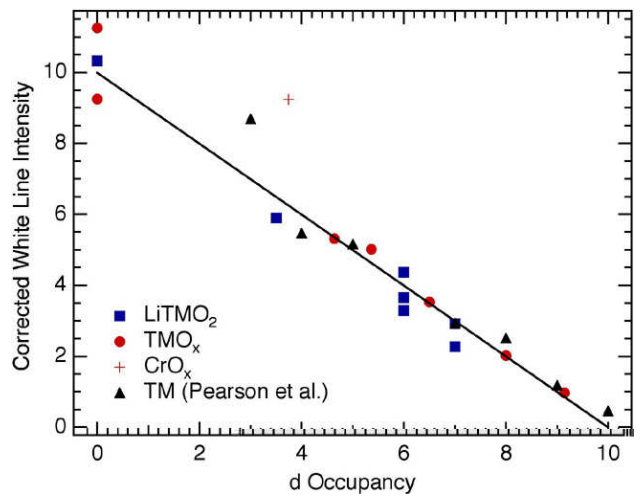}
\caption{(Color online) Normalized white-line intensity for a series of transition metals, transition-metal oxides, and lithium transition-metal oxides plotted
vs.~$d$ occupancy. The solid line represents the expected white line line intensity based upon equation \ref{eq4}.}
\label{ivsd}
\end{figure}

A plot of the relationships between the integrated white-line intensity and the $3d$ occupancy for the transition metals \cite{Pearson1}, transition-metal oxides,
and lithium transition-metal oxides are displayed in Figure~\ref{ivsd}.
The results of Pearson et al.~shown in Figure \ref{ivsd} have been corrected for the effects of the matrix elements. This figure clearly demonstrates a linear
relationship between the $d$ occupancy and the white-line intensity for the $3d$ transition metals. This relationship is consistent with equation
\ref{eq4}, which can be rewritten as:
\begin{equation}
\rho_{total}^u(3d) \approx I \frac{M_{\textup{continuum}}}{M_{\textup{white line}}},
\label{whitelinefit}
\end{equation}
where $\rho_{total}^u(3d)$ is the total number of unoccupied 3$d$ states. The data of Figure \ref {ivsd} (excluding the datum from CrO$_x$) have a standard
deviation from equation \ref{whitelinefit} of $\pm 0.028$. The white-line intensity from CrO$_x$ clearly deviates from the observed trend, most
likely due to interference from the overlapping oxygen
$K$-edge that could not be properly subtracted from the data. The small deviations from linearity in
Figure
\ref{ivsd} can be attributed to structural differences between the different transition-metal compounds. The white lines are sensitive to the local environments
of the transition-metal ion. However, the consistency of these results suggests that the white-line intensities are only weakly affected by the local structural
environments of the transition-metal ion. It is interesting to note that the results for the transition-metal oxides and lithiated transition-metal oxides are
consistent with those for the elemental metals \cite{Pearson1}. The consistency of these results suggest that the
$L_{2,3}$ white lines may be used to characterize the transition metal valence in novel compounds, and to quantify oxidation/reduction
reactions in transition-metal oxides.

\subsection{Charge Compensation in LiNi$_{0.8}$Co$_{0.2}$O$_2$}
The transition-metal $L_{2,3}$ white lines of Li$_{1-x}$Ni$_{0.8}$Co$_{0.2}$O$_2$ were analyzed to determine the role of the
different transition-metal ions in compensating for the charge of the intercalated Li$^+$ ion. The empirical relationship between white-line intensity and
oxidation state established in equation \ref{whitelinefit} is ideally suited to this problem. As previously mentioned, the white lines are sensitive to the local
environment of the transition metal. Therefore, measuring fractional differences in oxidation state between two structurally unique materials would be difficult.
However, in Li$_{1-x}$Ni$_{0.8}$Co$_{0.2}$O$_2$ the local structural environment of the transition metal is unchanged between $0<x<0.64$
\cite{Saadoune,Ronci,Chebiam}. Therefore, the white-line intensities should be sensitive enough to detect small changes in $d$ occupancy due to Li intercalation.

\begin{figure}
\includegraphics{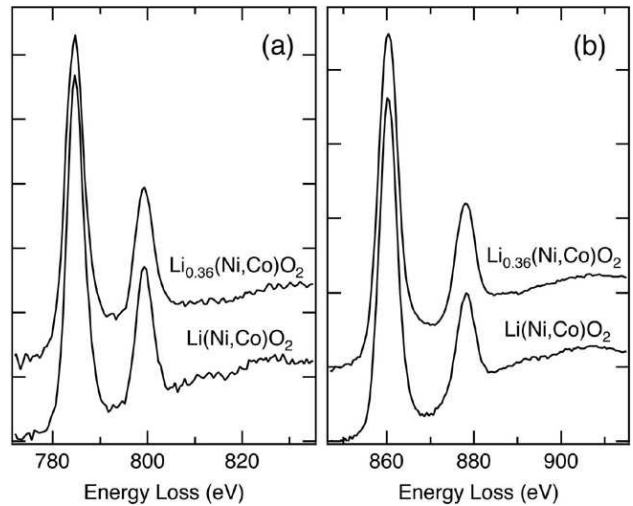}
\caption{({\bf a}) Co and ({\bf b}) Ni white lines in stoichiometric LiNi$_{0.8}$Co$_{0.2}$O$_2$ and delithiated
Li$_{0.36}$Ni$_{0.8}$Co$_{0.2}$O$_2$.}
\label{linicoo2_wl}
\end{figure}

\begin{figure}
\includegraphics{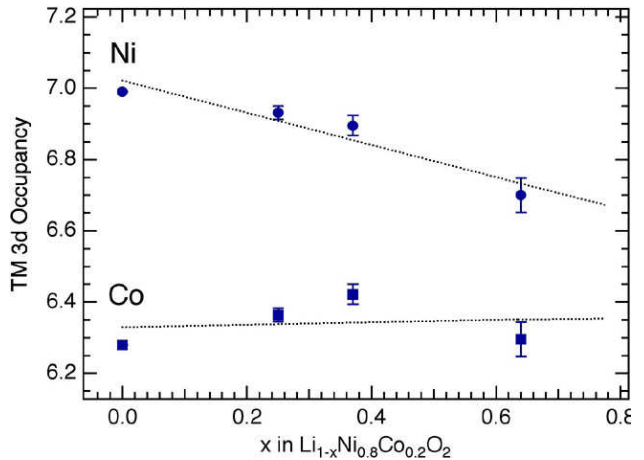}
\caption{(Color online) Nickel and cobalt $3d$ occupancies in Li$_{1-x}$Ni$_{0.8}$Co$_{0.2}$O$_2$.}
\label{oxidation_state}
\end{figure}

The cobalt and nickel white lines for stoichiometric LiNi$_{0.8}$Co$_{0.2}$O$_2$ and lithium-deficient Li$_{0.36}$Ni$_{0.8}$Co$_{0.2}$O$_2$ are
displayed in Figure \ref{linicoo2_wl}.
The cobalt white lines are invariant with lithium concentration, while the intensity of the
nickel white lines increases upon lithium extraction (Figure \ref{oxidation_state}).
Equation \ref{whitelinefit} was used to quantify the change in
$d$ occupancy. The error associated with this occupancy was determined from the white line intensities measured from the lithium transition-metal oxides (the
transition-metals and transition-metal oxides were not included). An error of approximately $\pm$0.4 electrons is associated with the absolute value of the
$d$ occupancy (not shown in Fig.\ref{oxidation_state}). However, the relative error in $d$ occupancy is small, and scales with the departure from
stoichiometry. The valence of the nickel ion increases by approximately 0.30 electrons in the range of lithium concentration $0.0\leq x\leq0.64$. Nickel accounts
for 80\% of the transition-metal ions, so the total charge compensation is approximately one quarter of an electron per transition metal. The remainder of the
charge is accommodated by the oxygen ion \cite{Graetzlinicoo2}.

\section{Conclusion}
A systematic study of the $L_{2,3}$ white lines has been performed on the $3d$ transition-metal oxides and lithiated transition-metal oxides. An obvious trend was
observed in the white-line ratio across the $3d$ series. The white-line ratio increases with $d$ occupancy when $n_d<5$ and decreases when $n_d>5$. The integrated intensity of the white lines
was also found to be sensitive to the oxidation state of the transition-metal ion. An empirical relationship between white-line intensity and $d$ occupancy was established for the
$3d$ transition-metal oxides and lithium transition-metal oxides. This relationship is a reliable standard for measuring $d$ occupancy in the $3d$ transition
metals. We have demonstrated the utility of this relationship by measuring the oxidation state of the nickel and cobalt ions in
Li$_{1-x}$Ni$_{0.8}$Co$_{0.2}$O$_2$. Upon delithiation, the nickel ion is partially oxidized (Ni$^{3.0+} \rightarrow$ Ni$^{3.3+}$), while the oxidation state of
the cobalt ion is unaffected.

\section{Acknowledgements}
This work was supported by the Department of Energy through Basic Energy Sciences Grant No.~DE-FG03-00ER15035.

\bibliography{bibliography}
\bibliographystyle{unsrt}

\end{document}